\documentclass{moriond}
\usepackage{xspace}
\usepackage{amsmath,amssymb}
\usepackage{environ}
\usepackage{hyperref}
\usepackage[displaymath, mathlines]{lineno} 

\def\ra{\rightarrow}

\def\be{\begin{equation}}
\def\ee{\end{equation}}
\def\bea{\begin{eqnarray}}
\def\eea{\end{eqnarray}}

\NewEnviron{comment}{}
\newcommand\con{\RenewEnviron{comment}{\color{red}[\BODY]}}

\con

\newcommand{\Photo}{}
\input{belle2-symbols}

\begin{document}

\vspace*{4cm}
\title{Time-dependent \CP violation results at \belletwo}

\author{ M. Veronesi, on behalf of the \belletwo Collaboration}

\address{Iowa State University, Ames, USA}

\maketitle\abstracts{
We report updates on time-dependent \CP-violation observables at \belletwo.
The benchmark measurements of the \Bz lifetime \taud and mixing frequency \dmd using flavor specific hadronic decays and the determination of the \CP-violating phase \sintpo in \ccs transitions have been performed using data collected between 2019--2021.
These analyses use only half of the current available dataset and are still statistically limited, showing the excellent performance of the detector and readiness of the analysis tools.
We present three new results on the effective value of \sintpo in \qqs transitions, which are highly sensitive to generic non-Standard Model (SM) physics amplitudes, using the full dataset collected between 2019--2022.
}

\section{Introduction}
Measurements of the \Bz mixing frequency \dmd with flavor-specific decays and the determination of the \CP-violating phase \sintpo in \ccs transitions are important elements to constrain the unitarity of the Cabibbo-Kobayashi-Maskawa (CKM) matrix in the SM.
On the other hand, measurements of time-dependent \CP-violation in \qqs transitions offer a powerful probe for generic new physics, as they proceed through loop-suppressed decays which are potentially affected by non-SM amplitudes~\cite{Beneke:2005pu}. 
However, this class of decays usually involves neutral particles in the final state, that are experimentally challenging to reconstruct.
This, combined with the small branching fractions, makes the current average of available measurements statistically less precise than the theory prediction.
\belletwo is in the unique position to improve the current experimental knowledge due to its capabilities with vertex determination and efficient reconstruction of neutral particles.

\belletwo~\cite{Abe:2010gxa} is a high-energy physics experiment at the SuperKEKB collider~\cite{Akai:2018mbz}, operating at the \FourS resonance.
The detector is designed to reconstruct the decays of heavy mesons and $\tau$ leptons in energy-asymmetric \epem collisions.
Of particular importance for the measurement of time-dependent observables is the innermost part of the detector, equipped with a two-layer silicon pixel detector (PXD), surrounded by a four-layer double-sided silicon-strip detector (SVD).
The dataset used for the analyses presented here was collected with only one sixth of the second PXD layer installed.
\BBbar events are produced in a quantum-entangled state from the decay of an \FourS resonance.
The proper-time difference \dt is estimated using the decay vertex positions of the two \B mesons in the event along the boost axis.
In spite of the lower boost compared to KEKB, the upgraded detector is able to achieve a better vertex resolution than its predecessor.
In addition, the knowledge of decay times is enhanced by the constraint from the beam spot profile in combination with the new nano-beam scheme, achieving a $\dt$ resolution of less than 1~\ps.

\section{Measurement of \taud, \dmd and \sintpo with 2019--2021 data}
\begin{figure}
\centering
\includegraphics[width=0.45\textwidth]{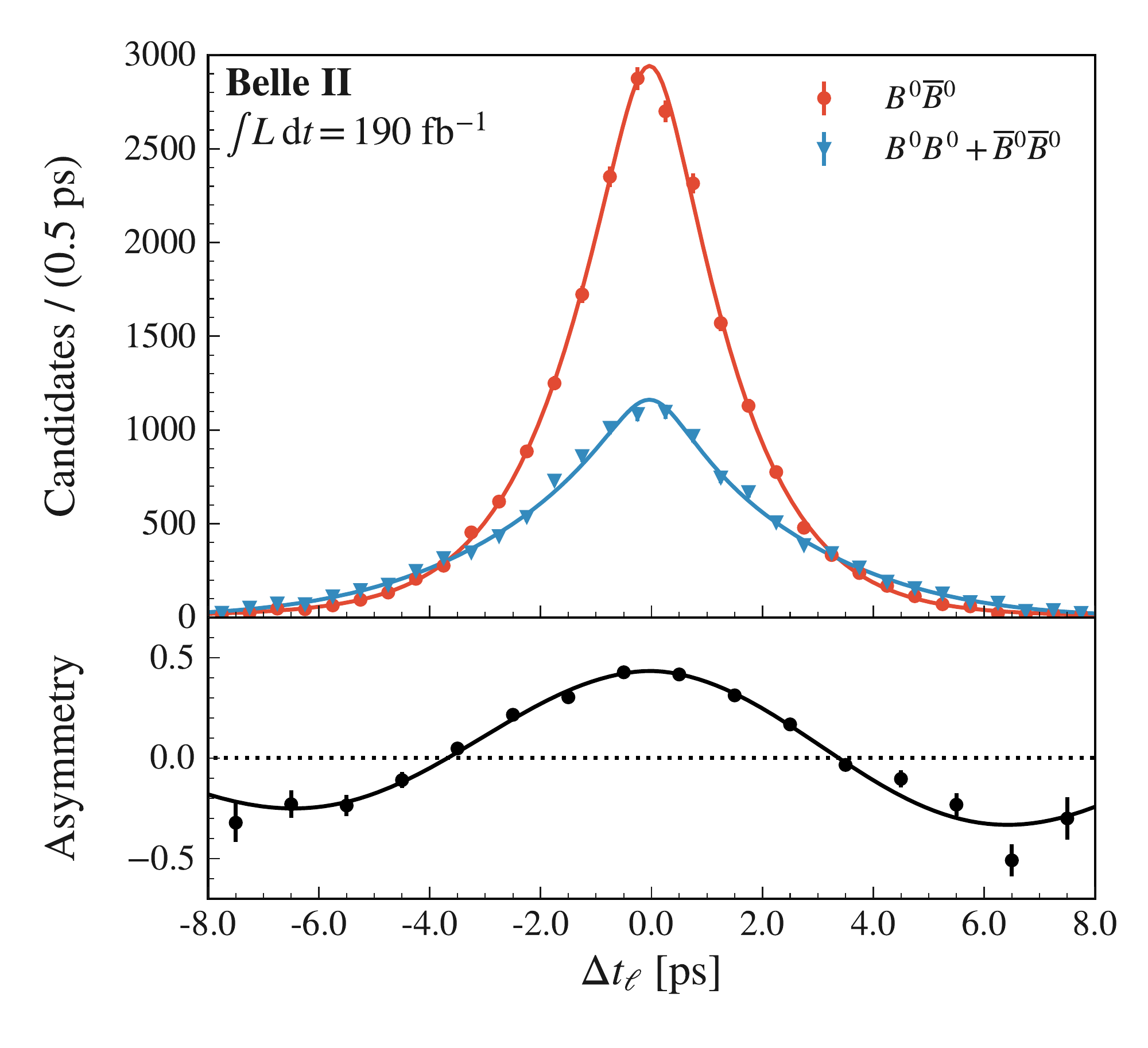}
\includegraphics[width=0.45\textwidth]{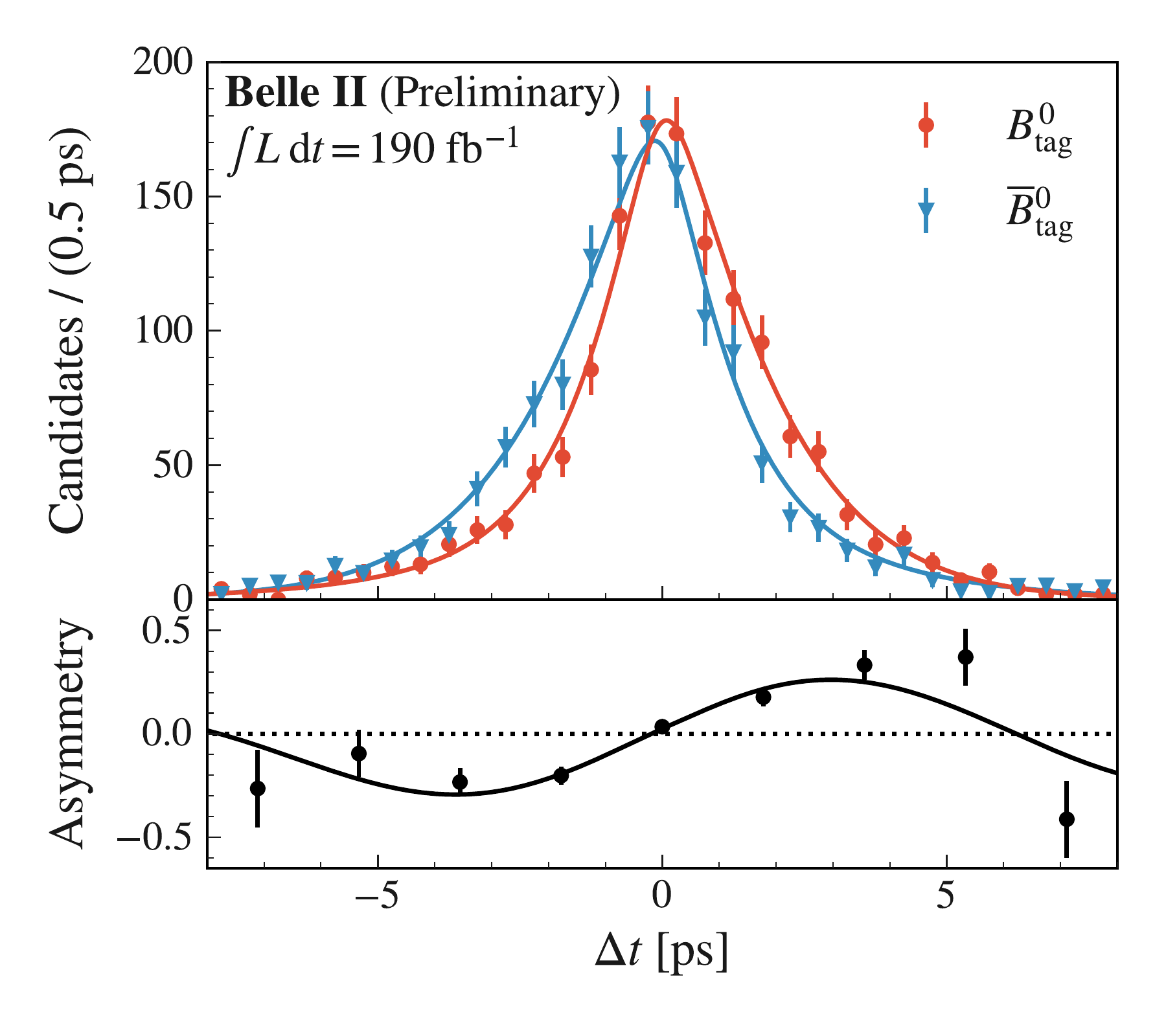}
\caption{Projections of the \dt fit on the \dpi  (left) and \jpsiks (right) samples.}
\label{fig:ccs-asymm}
\end{figure}

The distribution of the decay time difference \dt for flavor-specific \Bz decays is: 
\begin{equation}
\mathcal{P}(\dt,q)=\frac{e^{-|\dt|/\taud}}{4\taud} \Big\{ 1+q\cos(\dmd\dt) \Big\},
\label{eq:flav}
\end{equation}
where \taud is the \Bz lifetime, \dmd is the $\Bz-\Bzb$ mixing frequency, and $q$ is $+1$ ($-1$) when the two \B mesons have the opposite (same) flavor.
The flavor of the other \Bz is identified using a category-based \B-flavor tagging algorithm~\cite{Belle-II:2021zvj} from the inclusive properties of particles in the event that are not associated with the signal candidate.

The measurement of \taud and \dmd allows to test the QCD theory of strong interactions at low energy~\cite{Lenz:2014jha} and to constrain the side of the CKM triangle.
In addition, one is able to experimentally determine the \dt resolution function and flavor tagging parameters diluting the observable oscillations. 
These inputs are needed for the measurement of time-dependent \CP asymmetries in \Bz decays to \CP eigenstates, for which the \dt distribution is:
\begin{equation}
\mathcal{P}(\dt,q)=\frac{e^{-|\dt|/\taud}}{4\taud} \Big\{ 1+q\big[ \ACP \cos (\dmd \dt) + \SCP \sin(\dmd \dt)\big] \Big\},
\label{eq:dt_theo}
\end{equation}
where $q$ is the flavor of the other \Bz in the event ($q=+1$ for \Bz and $q=-1$ for \Bzb), and \ACP and \SCP are the direct and mixing induced \CP asymmetries.
For \jpsiks decays, the values of \ACP and \SCP are expected to be equal to zero and \sintpo, respectively, in the SM.

The most recent \belletwo analyses~\cite{Belle-II:2023nmj,Belle-II:2023bps} are based on a sample of 190\invfb collected at the \FourS center-of-mass energy and corresponding to $200\times10^6$ \BBbar pairs.
We reconstruct $33 317$ signal \dpi decays and $2755$ signal \jpsiks events.
The background-subtracted~\cite{Pivk:2004ty} \dt distributions and corresponding flavor specific and mixing induced \CP asymmetries are shown in Fig.~\ref{fig:ccs-asymm}.
The measured lifetime, mixing frequency and \CP asymmetries are reported in Tab.~\ref{tab:ccs} together with the world average values~\cite{HeavyFlavorAveragingGroup:2022wzx}.
For the lifetime and mixing measurements, the largest sources of systematic uncertainty are due to the resolution function parameters fixed from simulation and detector misalignment.
For the determination of the direct and mixing-induced \CP asymmetries, the dominant sources of systematic uncertainty are the tag-side interference (\ie~the presence of $\bquark \to \uquark\cquarkbar\dquark$ Cabibbo-suppressed decay
in the tagging \Bz) and the limited statistical knowledge of the flavor tagging and resolution parameters from the \dpi calibration sample.
Although not yet as precise as the current world-leading measurements, these results are still statistically limited and have systematic uncertainties comparable to those of previous generation \B-factories.

\begin{table}
\centering
\caption{
Comparison of recent \belletwo results (where the first uncertainties are statistical, while the second are systematic) and world average values of the \Bz lifetime, mixing frequency and \CP asymmetries in \ccs transitions.}
\label{tab:ccs}
\begin{tabular}{|lll|}
\hline
Observable & \belletwo ($190\invfb$) & World Average \\
\hline
\taud & $1.499 \pm 0.013 \pm 0.008\ps$      & $1.519 \pm 0.004\ps$ \\
\dmd  & $0.516 \pm 0.008 \pm 0.005\invps$   & $0.5065 \pm 0.0019\invps$ \\
\ACP(\ccs)  & $0.094 \pm 0.044^{+0.042}_{-0.017}$ & $0.005\pm0.015$ \\
\SCP(\ccs)  & $0.720 \pm 0.062 \pm 0.016$         & $0.699\pm0.017$ \\
\hline
\end{tabular}
\end{table}

\section{Measurement of \sintpo in \qqs transitions with 2019-2022 data}
\begin{figure}
\centering
\includegraphics[width=0.9\textwidth]{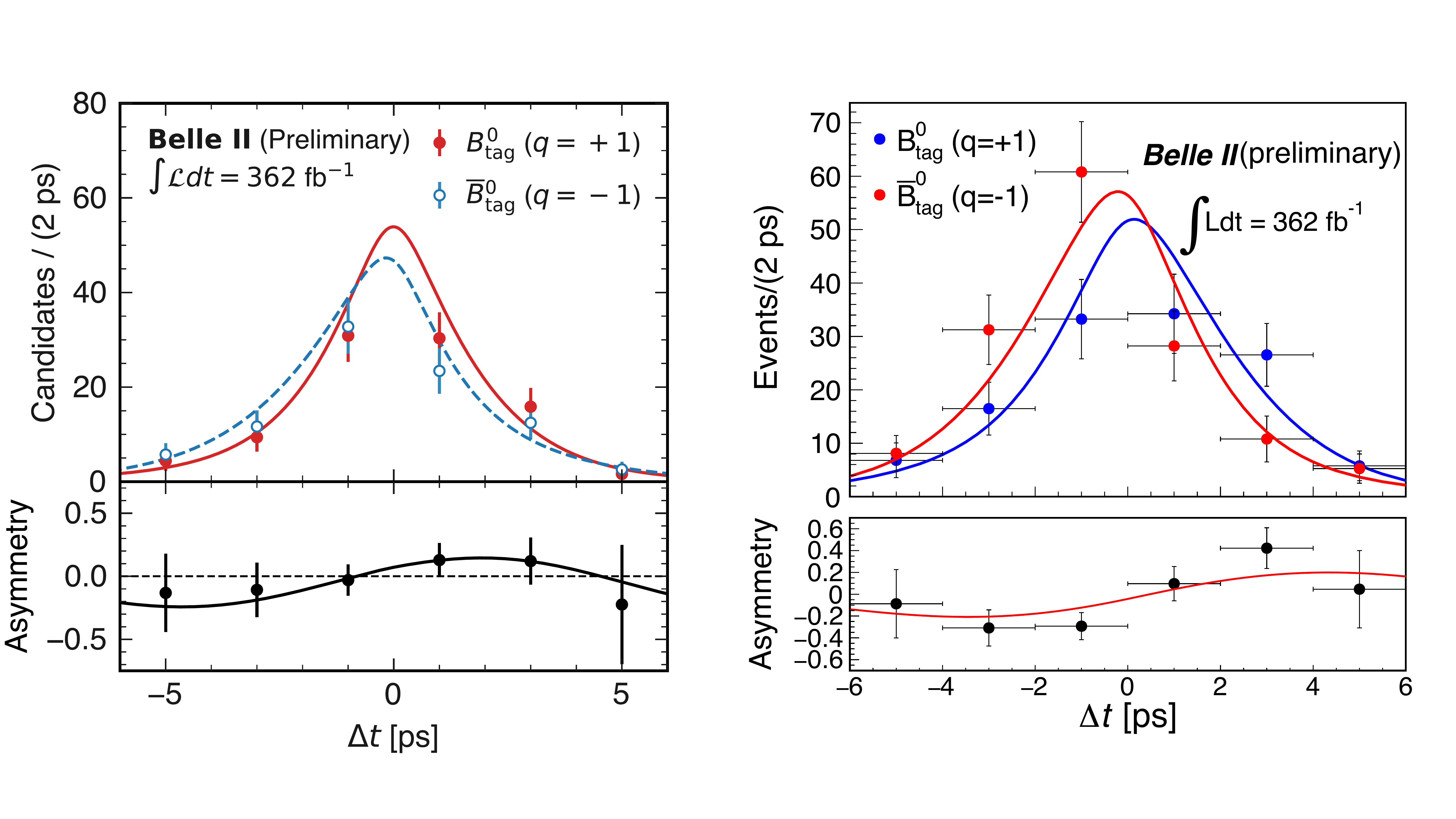}\\
\includegraphics[width=0.45\textwidth]{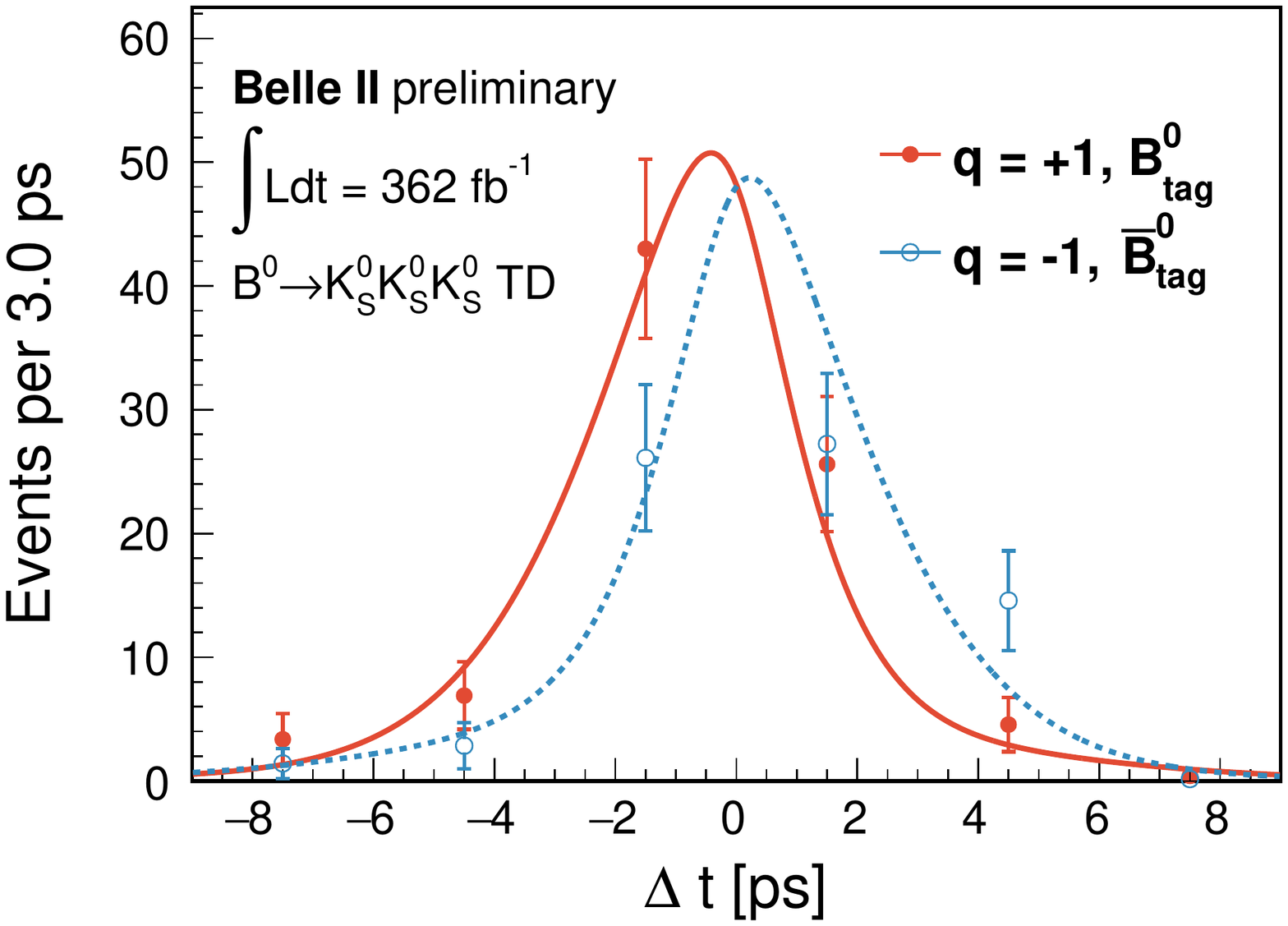}
\includegraphics[width=0.45\textwidth]{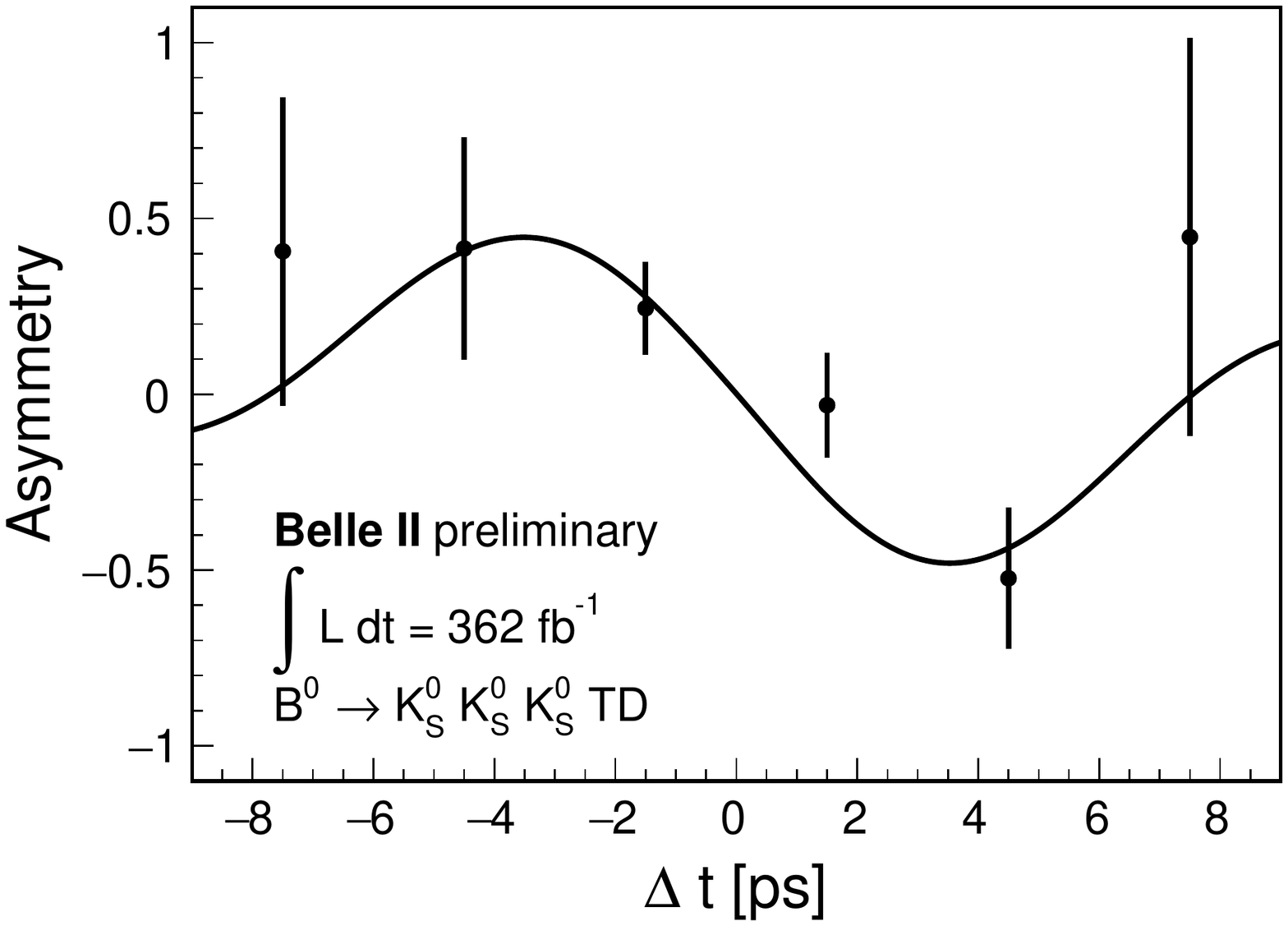}
\caption{Projections of the \dt fit and \CP asymmetries in \qqs decays: \phiks (top left), \kspiz (top right) and \ksksks (bottom).}
\label{fig:qqs-asymm}
\end{figure}

The decays \phiks, \ksksks and \kspiz all proceed through \qqs gluonic penguin transitions and therefore provide inputs to the effective value of \sintpo.
\belletwo has recently reported three new measurements using a sample of $362\invfb$, corresponding to $387\times 10^6$ \BBbar pairs.
The three analyses adopt similar techniques to separate signal from background, \eg multi-dimensional likelihood fits the beam-constrained mass \mbc, energy difference \deltae and transformed output of the classifier \ocs combining several continuum suppression variables.
In addition, they use the flavor tagging and,  in the case of \phiks, resolution function parameters from the \dpi calibration sample.
The background-subtracted~\cite{Pivk:2004ty} \dt distributions are displayed in Fig.~\ref{fig:qqs-asymm} and the measured \CP asymmetries are reported in Tab.~\ref{tab:qqs}.

\subsection{\phiks}
The \phiks decay vertex is reconstructed from the two prompt tracks of the \phikk decay, therefore, it has a similar \dt resolution as the \jpsiks mode.
In addition to the dominant continuum \qqbar background, it suffers from a sizeable contribution from non-resonant \nonres decays with the same final state but opposite \CP eigenvalue, diluting the observable \CP asymmetries.
In order to disentangle the non-resonant background component, we perform a multidimensional fit including the cosine of the helicity angle, in which the \phiks and \nonres have different distributions.
In total, we reconstruct $162\pm17$ signal \phiks and $21\pm12$ background \nonres events.
We estimate the residual effect of neglecting interference using a MC sample generated with a complete Dalitz description of the decay.
The analysis is validated on generic MC and on the \phikp control channel in data, which features similar backgrounds, vertexing and null \CP asymmetries.
The statistical sensitivity on \ACP is on par with the world's best measurements.
When compared to the \belle~\cite{PhysRevLett.98.031802} and \babar~\cite{BaBar:2005xng} analyses using a similar quasi-two body strategy, there is a 10 to 20\% statistical improvement on \SCP for the same number of signal events.
The dominant sources of systematic uncertainty stem from the bias induced by the fit model used to disentangle signal from backgrounds and neglecting the contribution from additional mis-reconstruced \BBbar backgrounds in the fit.

\subsection{\ksksks}
The \ksksks decay proceeds through the same underlying \sss quark transition as of \phiks.
It has the advantage of not being affected from opposite-\CP backgrounds.
However, since \KS decay on average outside of the pixel detector, it is experimentally challenging due to the absence of prompt tracks to form a vertex.
The decay vertex reconstruction relies on the \KS trajectory and profile of the interaction point.
In order to achieve the best statistical sensitivity, the dataset is divided into ``time-differential'' (TD) events, for which the \KS carry sufficient information from the vertex detector, and ``time-integrated'' (TI) events, for which the decay vertex is poorly constrained.
The TD events are used in the time-dependent \CP fit, while TI events are used only to measure \ACP.
In addition, the resolution function parameters obtained in simulation are scaled in data by including the \kskskp control channel in the combined fit.
In total, we reconstruct $158^{+14}_{-13}$ TD and $62\pm9$ TI events.
The statistical sensitivity on \ACP is on par with the world's best measurements.
The leading sources of systematic uncertainty are the bias induced by the fit model and calibration of the flavor tagging.

\subsection{\kspiz}
The \kspiz decay belongs to the same class of \qqs decays as \phiks and \ksksks.
It has a higher effective branching fraction than \phiks and \ksksks but slightly larger theoretical uncertainties~\cite{Beneke:2005pu}.
The signal reconstruction requires excellent performance with neutrals, due to the absence of prompt tracks and presence of a \piz in the final state.
The analysis follows a similar strategy as \ksksks, dividing the dataset into TD and TI events to retain the information on \ACP from events with poor \dt resolution.
In total, we reconstruct $415^{+26}_{-25}$ signal events.
The analysis strategy is validated on \jpsiks data, reconstructed without the vertex information from the \jpsi.
The statistical sensitivity on \ACP and \SCP is already on par with the world's best determinations in spite of the smaller dataset.
The dominant contribution to the systematic uncertainty arise from neglecting possible \CP asymmetries in the backgrounds and from the calibration of the resolution function.

\begin{table}
\caption{Comparison of recent \belletwo results (where the first uncertainties are statistical, while the second are systematic) and world average of \CP asymmetries in \qqs transitions.}
\label{tab:qqs}
\renewcommand{\arraystretch}{1.3}
\centering
\begin{tabular}{|llll|}
\hline
\multicolumn{2}{|l}{Observable} & \belletwo ($362\invfb$) & World Average \\
\hline
\phiks  & \ACP & $0.31 \pm 0.20^{+0.05}_{-0.06}$ & $-0.01\pm0.14$ \\
        & \SCP & $0.54 \pm 0.26^{+0.06}_{-0.08}$ & $0.74^{+0.11}_{-0.13}$ \\
\ksksks & \ACP & $0.07^{+0.15}_{-0.20} \pm 0.02$ & $0.15\pm0.12$ \\
        & \SCP & $-1.37^{+0.35}_{-0.45}\pm 0.03$ & $-0.83\pm0.17$ \\
\kspiz  & \ACP & $0.04^{+0.15}_{-0.14}\pm 0.05$        & $-0.01\pm0.10$ \\
        & \SCP & $0.75^{+.20}_{-0.23}\pm 0.04$  & $0.57\pm0.17$ \\
\hline
\end{tabular}
\end{table}

\section{Summary}
\belletwo has performed measurements of the \Bz lifetime and mixing frequency with flavor-specific decays and \CP asymmetries in \ccs transitions using half of its dataset.
These high-yield analyses require the accurate modeling of the vertex resolution and flavor tagging response, which represent important milestones in the development of time-dependent analyses.
In addition, we report recent results on \CP violation in \qqs transitions using the full \belletwo datasets, where some observables are already competitive with the world's most precise measurements, albeit using much less luminosity.
Due to its excellent neutral reconstruction capabilities, \belletwo is in the unique position to improve our current experimental knowledge on these modes, that are essential to probe generic non-SM physics in loops.


\section*{References}

\bibliography{references}



\end{document}